\documentclass{article}
\usepackage[utf8]{inputenc}
\usepackage{pgfplots}
\usetikzlibrary{patterns}
\pgfplotsset{compat=1.17}
        \usepackage{setspace}

\newcommand{\rev}[1]{{\color{black}#1}}

\newcommand{\plosfootnote}[1]{\footnote{#1}}

\newcommand{\plosbibstyle}{alphabetic}
\title{The Role of Author Identities in Peer Review}

\usepackage{amsthm,amsmath, amssymb, amsfonts}
\usepackage{hyperref}
\usepackage[margin=1in]{geometry}
\usepackage{graphicx}
\usepackage{url}            
\usepackage{caption,subcaption}
\usepackage{enumitem}
\setlist[enumerate]{leftmargin=2em}
\setlist[itemize]{leftmargin=2em}

\usepackage{multirow}

\usepackage[style=\plosbibstyle,backend=bibtex,giveninits=true,doi=false,isbn=false,url=false,eprint=false,maxbibnames=99]{biblatex}
\AtEveryBibitem{%
\clearfield{pages}%
  \clearfield{volume}%
  \clearfield{number}%
  \clearfield{month}%
}

\author{}
\date{} 

\begin{document}

\maketitle
\vspace{-1.4cm}

\begin{center}
\begin{tabular}{c}
{\large Nihar B. Shah}\\
Carnegie Mellon University\\
{\tt nihars@cs.cmu.edu} \\
\end{tabular}
\end{center}

~\\

\begin{abstract}
There is widespread debate on whether to anonymize author identities in peer review. The key argument for anonymization is to mitigate bias, whereas arguments against anonymization posit various uses of author identities in the review process. The Innovations in Theoretical Computer Science (ITCS) 2023 conference adopted a middle ground by initially anonymizing the author identities from reviewers, revealing them after the reviewer had submitted their initial reviews, and allowing the reviewer to change their review subsequently. We present an analysis of the reviews pertaining to the identification and use of author identities. Our key findings are: (I) A majority of reviewers self-report not knowing and being unable to guess the authors' identities for the papers they were reviewing. (II) After the initial submission of reviews, 7.1\% of reviews changed their overall merit score and  3.8\% changed their self-reported reviewer expertise. (III) \rev{There is a very weak and statistically insignificant correlation of the rank of authors' affiliations with the change in overall merit; there is a weak but statistically significant correlation with respect to change in reviewer expertise.} We also conducted an anonymous survey to obtain opinions from reviewers and authors. The main findings from the 200 survey responses are: (i) A vast majority of participants favor anonymizing author identities in some form. (ii) The ``middle-ground'' initiative of ITCS 2023 was appreciated. (iii) Detecting conflicts of interest is a challenge that needs to be addressed if author identities are anonymized.  Overall, these findings support anonymization of author identities in some form (e.g., as was done in ITCS 2023), as long as there is a robust and efficient way to check conflicts of interest.
\end{abstract}

\section{Introduction}
\label{SecIntro} 

There is a lot of debate in various research communities on whether to anonymize author identities or not. The primary argument for anonymization is that it mitigates bias in the review process. Indeed, a number of experiments in computer science~\cite{tomkins2017reviewer,manzoor2020uncovering} and in other fields~\cite{okike2016single,blank1991effects,ross2006effect,garfunkel1994effect,fisher1994effects,huber2022nobel} have found evidence that reviews are biased if author identities are visible to the reviewers. The most consistent finding across these studies is that reviews are biased favorably towards authors from higher-ranked affiliations or authors who are famous. 

Traditionally, peer review in the field of theoretical computer science does \emph{not} anonymize author identities. This topic has been hotly debated in the field with various opinions and discussions~\cite{Fortnow2018,Barak_2018,Venkatasubramanian_2018,Reingold_2018,Mitzenmacher_2018}. Some of the stated reasons against anonymization pertain to using author identities to gain more confidence in hard-to-verify mathematical proofs or wacky ideas, more scrutiny for papers by authors with a history of bad science, ensuring similar papers are from different authors, mitigating incremental work by same authors, and allocate reviewers' limited time and effort accordingly. 

This study is focused on the Innovations in Theoretical Computer Science (ITCS) 2023 conference. In computer science, conferences typically review full papers (and not just abstracts), are the terminal venue of publication, are rated at least at par with journals, and are usually competitive with 15-25\% acceptance rates.

Given the aforementioned debate regarding anonymization of authors, in the review process of the ITCS 2023 conference, we took the following middle-ground approach. Author identities were initially hidden from reviewers. Once the reviewer submitted their review, they could see author identities. Reviewers were subsequently allowed to modify their review if they wished. The rationale behind this approach was that it could mitigate bias at least in the initial impression, and later allow reviewers to use author identities in a manner that has been posited in aforementioned discussions. Note that the authors were free to upload their (non anonymous) papers to preprint servers, social media, etc.

In this paper we address three key questions pertaining to author identities in the review process:
\begin{itemize}
\item \emph{Identification:} Given that author identities were initially hidden, how many reviewers could identify the authors?
\begin{itemize}
    \item We address this question by asking reviewers to self-report their knowledge of author identities when they submit their review (Section~\ref{SecBasicReview}). 
\end{itemize}
\item \emph{Usage:} Given that author identities were made visible to reviewers after they submitted initial reviews, how did the reviewers subsequently use the author identities to modify their reviews?
\begin{itemize}
    \item We address this question by analyzing the change in the reviews after they are initially submitted (Section~\ref{SecChangeAnalysis}). 
\end{itemize}
\item \emph{Opinions:} What opinions or prespectives do participants have regarding anonymization of author identities? 
\begin{itemize}
    \item We address this question via an anonymous survey of the participants in the ITCS 2023 review process (Section~\ref{SecSurvey}). 
\end{itemize}
\end{itemize}
We envisage these results to be useful in designing author-anonymization policies in an evidence-based fashion.


\section{Related work}

A number of studies investigate whether revealing identities of authors to reviewers results in biases in the reviews. Perhaps the most well-known such experiment in computer science involves a randomized controlled trial in the Web Search and Data Mining (WSDM) 2017 conference~\cite{tomkins2017reviewer} in data mining. This experiment found biases favoring authors who were famous or from top-ranked institutions. (See the paper~\cite{stelmakh2019testing} for some concerns regarding the experimental design and methods employed in~\cite{tomkins2017reviewer}.) The study~\cite{manzoor2020uncovering} conducted an observational study that found biases favoring authors from top-ranked affiliations in the International Conference on Learning Representations (ICLR) conference in machine learning. A number of studies in other fields also find biases favoring authors who are famous or from top-ranked institutions~\cite{okike2016single,blank1991effects,ross2006effect,garfunkel1994effect,fisher1994effects,huber2022nobel}; \rev{see~\cite{haffar2019peer} for an overview on biases in peer review}.

The studies~\cite{ware2008peer,mulligan2013peer} survey researchers about their perceptions of the peer-review process. The surveys reveal that researchers across a number of fields prefer anonymizing author identities. In the adjoining field of information theory where anonymization is currently not the norm, a survey of the members of the community~\cite{shah2021itsocsurvey} shows significant support for trying out anonymization.

\rev{In addition to mitigating bias, another aspect of reviews is their ``quality''. A randomized controlled trial between anonymized and non-anonymized reviewing found that there was no difference in the quality of reviews as assessed by editors and authors~\cite{justice1998does}.}

The problem of measuring possible bias in peer review has also spurred research on designing and evaluating the methods to do so. The aforementioned work~\cite{stelmakh2019testing} provides an experimental design framework, statistical tests, and associated theoretical guarantees for conducting such a measurement in a randomized controlled trial. The paper~\cite{manzoor2020uncovering} uses difference-in-difference techniques to estimate bias exploiting the fact that the ICLR conference changed from not anonymizing to anonymizing author names in a certain year. In addition, the paper~\cite{manzoor2020uncovering} also develops methods to test for biases based on the review text rather than just the review scores. The pair of papers~\cite{madden2006impact,snodgrass2006single} debate the methods of computing the bias, with different metrics leading to different conclusions. Finally, the paper~\cite{jecmen2021split} focuses on the tradeoff between running such a controlled experiment in the peer-review process and the efficiency of the peer-review process. 

Even when author identities are not included in the submissions, there are ways in which they may be deciphered. First, a reviewer may have seen the paper outside of the review process such as on a preprint server, on social media, or in a talk. For instance, a little over half of the papers submitted to the NeurIPS 2019 conference were posted on arXiv, and among these papers, 21\% were seen by at least one reviewer~\cite{beygelzimer2019neurips}. Second, some reviewers may actively search for their assigned papers online. For instance, the study~\cite{rastogi2022arxiv} found that over a third of the reviewers in two top conferences searched for their assigned paper online. Third, the reviewer may be able to guess the identity of the authors based on the content of the paper. The study~\cite{goues2018effectiveness} asked reviewers in three conferences which anonymized author identities to guess the authors of papers.  
A total of 70\%-86\% of the reviews did not provide any guess (which could mean that the reviewer did not have any idea of the identity or that the reviewer chose to just not answer the question). Among the subset of reviews which contained a guess, 72\%-85\% guessed at least one author correctly. Some other works~\cite{shawndra03identification,caragea2019myth,matsubara2020citations} investigate the possibility of identifying authors based on content by designing machine learning algorithms to predict the authors based on the paper's content. In our experiment, we also investigate this aspect in ITCS 2023 by asking reviewers to self-report whether they can identify the authors. \rev{The paper~\cite{cho1998masking} aims to identify the factors that are associated with successful anonymization in seven biomedical journals. They found no association with whether the journal had a long-standing policy of anonymized reviewing, but they did find that anonymization was less successful when the reviewer had a greater research experience. }

Opening up author identities to reviewers near the end of the review process has been employed before in other computer science conferences~\cite{Hicks_2012}. There, the author identities are usually revealed right before the program committee meeting. Their stated reasons for this approach are that fully anonymous review might penalize a paper for failing to cite a certain prior work but this prior work may be unpublished work by the same authors, and anonymous review may not discover certain conflicts which the reviewer may know about. Our middle ground approach is also motivated by various other frequently mentioned uses of author identities (Section~\ref{SecIntro}). Most importantly, a key objective of this approach in ITCS 2023 was to understand the use of author identities by reviewers to enable subsequent policy-makers to design policies using this evidence.

\rev{An alternative policy that lies between non-anonymization and compulsory anonymization is that of not making anonymization compulsory, but allowing the authors the option to anonymize their papers. The paper~\cite{Smirnova_Romero_Teplitskiy_2023} studies such a policy in a number of journals from IOP publishers, finding that authors of 22\% of papers volunteered to anonymize, and furthermore, the highest-prestige authors anonymized less often but still substantially. This policy increased the acceptance rate of papers by ``low-prestige'' authors by 5.6\%. Such a policy in the Nature group of journals had resulted in 12\% of papers being anonymized for reviewing~\cite{nature_anon_uptake_2017}.
}

Questions about bias might arise more when decisions can be subjective, for instance, relying on criteria such as predictions of potential impact or novelty or interestingness to rank papers. There are various experiments in computer science and elsewhere that have found high disagreement among reviewers in terms of which paper is better~\cite{obrecht2007examining, fogelholm2012panel,nips14experiment,pier2017your,beygelzimer2021neuripsconsistency}, and more recently, experiments that have also found high disagreements among co-authors about their jointly authored papers~\cite{rastogi2022authors}. Consequently, there have been various debates about the low acceptance rates and ``selectivity'' in computer science conferences~\cite{fortnow2009viewpoint,parhami2016low,vardi2017divination}. 

Our work focuses on biases associated with author identities. In addition to this, there are a variety of other issues in peer review. See~\cite{shah2022surveyextended} for details. These issues lead to a variety of challenging theoretical problems which can lead to considerable practical impact if solved, but are not well explored in theoretical computer science.


\section{Main results 1: Analysis of reviews} 
In this section, we analyze the reviews to understand their dependence on author identities. We begin with a brief overview and statistics of the review process at ITCS 2023, and then present our analysis of the change in the reviews after the reviewer could see the author identities.\plosfootnote{See \url{https://aspredicted.org/ng77i.pdf} for the preregistration of this analysis.}

\subsection{Overview and basic statistics of the review process}
\label{SecBasicReview}

The ITCS 2023 conference received a total of 235 paper submissions. 
The set of reviewers comprised a program committee (PC) and external reviewers. The 40-member program committee were ultimately responsible for the acceptance/rejection decisions on all papers.  The program committee also included the program chair of the conference. The members of the program committee could review papers themselves or ask experts not in the program committee, called external reviewers, to review. In ITCS 2023, there were 40 members in the program committee who did a total of 405 reviews. There were 315 external reviewers who did 361 reviews. 

Participants were notified of the experiment before the review process started and were provided an option to decline participation in the experiment. (None of the participants declined). This work was reviewed and approved by the Carnegie Mellon University Institutional Review Board (IRB). The analysis pertained to reviews already provided in the conference and the IRB waived informed consent.

The conference asked authors to anonymize their identities on the submitted paper. The conference policy did not prohibit authors from posting their (non anonymous) papers online or on social media or give talks. Reviewers were not shown the authors' identities until they submitted a review. After a reviewer submitted their review, they could see the authors' identities for that paper, and could modify their review in any manner they chose. The reviewers were told of this policy beforehand, and were also told that the (changes in the) reviews will be analyzed.  

Reviewers were asked to fill out following form when submitting their review for any paper:

\noindent\rule{\textwidth}{1pt}
\begin{itemize} 
\item Overall merit: 5. Strong accept, 4. Accept, 3. Weak accept, 2. Weak reject, 1. Reject
\item Reviewer expertise: 4. Expert, 3. Knowledgeable, 2. Some familiarity, 1. No familiarity
\item Paper summary $\_\_\_\_\_\_\_\_\_\_\_\_\_\_\_\_\_\_\_\_\_\_\_\_\_\_\_\_\_\_\_\_\_\_\_\_\_\_\_\_\_\_$
\item Comments for authors $\_\_\_\_\_\_\_\_\_\_\_\_\_\_\_\_\_\_\_\_\_\_\_\_\_\_\_\_\_\_\_\_\_\_\_\_\_\_\_\_\_\_$
\item Comments for PC $\_\_\_\_\_\_\_\_\_\_\_\_\_\_\_\_\_\_\_\_\_\_\_\_\_\_\_\_\_\_\_\_\_\_\_\_\_\_\_\_\_\_$
\item Author identity: Please indicate your knowledge of author identities. [Visible to program chairs. Not visible to authors at any time.]
    \begin{itemize}
        \item[A.] I know the author identities (e.g., saw a talk or on arxiv etc.)
        \item[B.] I have a reasonable guess for author identities
        \item[C.] I have no idea about author identities
    \end{itemize}
\end{itemize}
\noindent\rule{\textwidth}{1pt}

\begin{figure*}
\centering
\hspace{-1.2cm}
\begin{subfigure}{.24\textwidth}
\begin{tikzpicture}
\begin{axis} [xtick={0, 1,2,3,4},
        width=100pt,
        height=2in,
        ylabel = {\% Reviews},
        ymin=0, ymax=50,
         nodes near coords, 
        xmin=-0.5,xmax=4.5,
        xticklabels={1,2,3,4,5},
        xticklabel style   = {align=center},
        every axis plot/.append style={
          ybar,
          bar width=10pt,
          bar shift=0pt,
          fill
        },
        ymajorgrids=true,
    ]
\addplot[fill = blue] coordinates{
    (0,9)
};
\addplot[fill = blue] coordinates{
    (1,28)
};
\addplot[fill = blue] coordinates {
    (2,31)
};
\addplot [fill=blue] coordinates {
    (3,29)
};
\addplot [fill=blue] coordinates {
    (4,3)
};
\end{axis}
\end{tikzpicture}
\caption{Overall merit}
\end{subfigure}
\hspace{-.3cm}
\begin{subfigure}{.24\textwidth}
\begin{tikzpicture}
\begin{axis} [xtick={0, 1,2,3},
        width=90pt,
        height=2in,
        ylabel = {\% Reviews},
        ymin=0, ymax=50,
         nodes near coords, 
        xmin=-0.5,xmax=3.5,
        xticklabels={1,2,3,4},
        xticklabel style   = {align=center},
        every axis plot/.append style={
          ybar,
          bar width=10pt,
          bar shift=0pt,
          fill
        },
        ymajorgrids=true,
    ]
\addplot[fill = yellow] coordinates{
    (0,6)
};
\addplot[fill = yellow] coordinates{
    (1,31)
};
\addplot [fill=yellow] coordinates {
    (2,44)
};
\addplot [fill=yellow] coordinates {
    (3,19)
};
\end{axis}
\end{tikzpicture}
\caption{Reviewer expertise}
\end{subfigure}
\hspace{-.5cm}
\begin{subfigure}{.45\textwidth}
\begin{tikzpicture}
\begin{axis} [xtick={0, 17, 33, 50, 66, 83, 100},
        width=3.5in,
        height=2in,
        ylabel = {\# Reviews},
        ymin=0, ymax=50,
        xmin=-0.5,xmax=100.5,
        xticklabels={0,500,1000,1500,2000,2500,3000},
        xticklabel style   = {align=center},
        every axis plot/.append style={
          ybar,
          bar width=1pt,
          bar shift=0pt,
          fill
        },
        ymajorgrids=true,
    ]
\addplot[fill = black] coordinates{
(0,7)(1,10)(2,18)(3,25)(4,19)(5,18)(6,37)(7,28)(8,21)(9,24)(10,43)(11,31)(12,45)(13,38)(14,39)(15,39)(16,28)(17,27)(18,30)(19,21)(20,15)(21,22)(22,16)(23,16)(24,12)(25,14)(26,10)(27,10)(28,5)(29,17)(30,8)(31,7)(32,2)(33,5)(34,1)(35,7)(36,1)(37,1)(38,3)(39,2)(40,2)(41,3)(42,2)(43,1)(44,2)(45,1)(47,2)(49,2)(50,3)(52,1)(53,1)(54,2)(98,1)(99,1)
};
\end{axis}
\end{tikzpicture}
\caption{Review length (number of words)}
\end{subfigure}
\caption{Basic statistics of the reviews in ITCS 2023.\label{FigReviewStatistics}} 
\end{figure*}
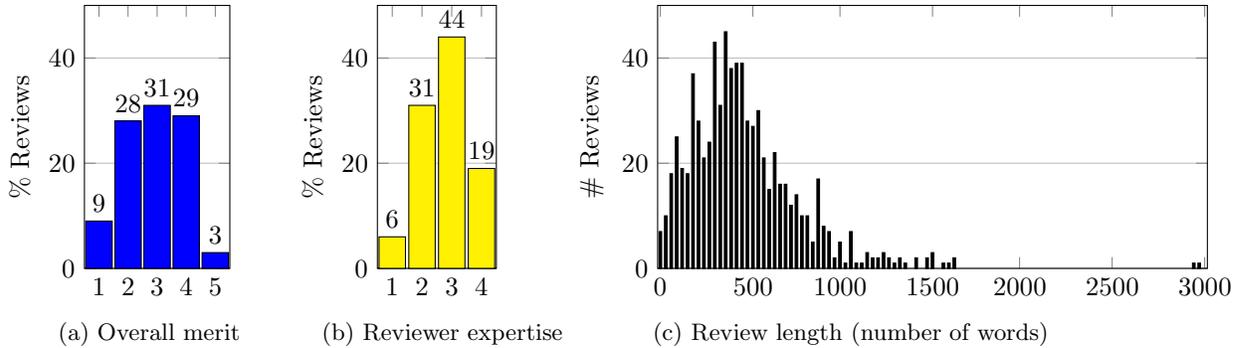

In Figure~\ref{FigReviewStatistics}, we plot the basic statistics pertaining to the reviews -- distribution of the overall merit scores, the self-reported reviewer expertise, and a histogram of the review length. Note that the reviewers could edit their reviews after they submitted an initial version, and these statistics pertain to the final state of each review. The mean length of the reviews was 480 words and median was 425 words. 

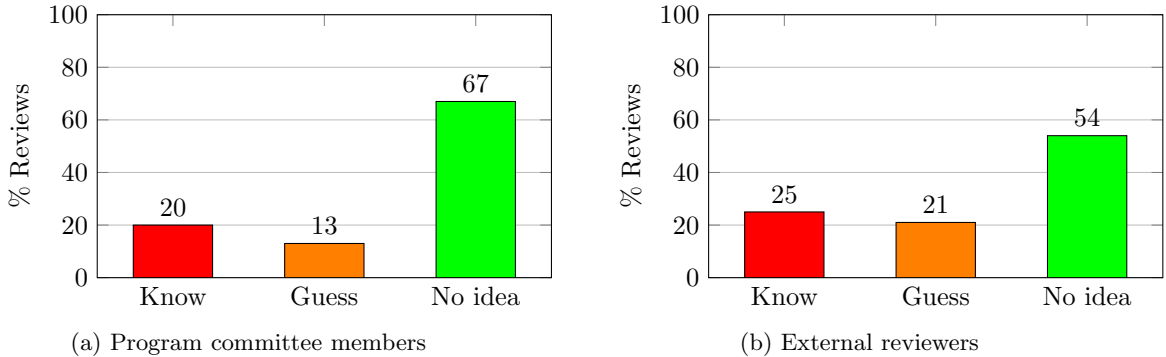
\begin{figure*}
\centering
\begin{subfigure}{.4\textwidth}
\begin{tikzpicture}
\begin{axis} [xtick={0, 1,2},
        width=3in,
        height=2in,
        ylabel = {\% Reviews},
        ymin=0, ymax=100,
         nodes near coords, 
        xmin=-0.5,xmax=2.5,
        xticklabels={Know,Guess,No idea},
        xticklabel style   = {align=center},
        every axis plot/.append style={
          ybar,
          bar width=30pt,
          bar shift=0pt,
          fill
        },
        ymajorgrids=true,
    ]
\addplot[fill = red] coordinates{
    (0,20)
};
\addplot[fill = orange] coordinates{
    (1,13)
};
\addplot[fill = green] coordinates {
    (2,67)
};
\end{axis}
\end{tikzpicture}
\caption{Program committee members}
\end{subfigure}
\qquad\qquad
\begin{subfigure}{.4\textwidth}
\begin{tikzpicture}
\begin{axis} [xtick={0, 1,2},
        width=3in,
        height=2in,
        ylabel = {\% Reviews},
        ymin=0, ymax=100,
         nodes near coords, 
        xmin=-0.5,xmax=2.5,
        xticklabels={Know,Guess,No idea},
        xticklabel style   = {align=center},
        every axis plot/.append style={
          ybar,
          bar width=30pt,
          bar shift=0pt,
          fill
        },
        ymajorgrids=true,
    ]
\addplot[fill = red] coordinates{
    (0,25)
};
\addplot[fill = orange] coordinates{
    (1,21)
};
\addplot[fill = green] coordinates {
    (2,54)
};
\end{axis}
\end{tikzpicture}
\caption{External reviewers}
\end{subfigure}
\caption{Distribution of reviewers' self reports about their knowledge of author identities.\label{FigAutIde}} 
\end{figure*}

We now come to the question about knowledge of author identities. It is important to note that this question was asked at the time the reviewer was submitting the review. This allows their answer to be based on a careful reading of the paper, and not just the title or abstract or a brief glance at the paper. In Figure~\ref{FigAutIde} we plot the distribution of reviewers' answers to this question. We observe among both program committee members and external reviewers, {\bf a majority of respondents report having no idea of the identities of the authors}.  

\subsection{Change in reviews after initial submission} 
\label{SecChangeAnalysis}

Recall that the reviewers could not see the identities of the authors of the paper initially, but the identities were made visible to them after they submitted a review for the paper. The reviewers could then edit their reviews, and in this section, we analyze the changes to the reviews after initial submission. 

Before delving into the details, we must address one confounder. Once a program committee member submitted their review, they could also see all other reviews submitted for that paper. Thus any change in their review may be due to other reviews, and such an influence of other reviews is well documented in the literature~\cite{teplitskiyasocial}. Consequently, for program committee members, we restrict attention to only those reviews that were the first to be submitted for the paper. In other words, in this analysis, we only consider a review by a program committee member if there was no other review submitted for that paper when this review was first submitted.  The external reviewers could not see other reviews at any time, and hence we consider all external reviews. This leaves us with 90 reviews by program committee members, along with the 361 reviews by external reviewers for our analysis. In the remainder of this section, we restrict attention to these 451 reviews. 

We tabulate the percentage and number of reviews that changed in Table~\ref{TabNumChanged}. In what follows, we provide a deeper analysis, where we also compute the correlation of the change in overall merit or reviewer expertise with the rank of the affiliations of the authors of the paper. We compute the rank of any affiliation based on \url{csrankings.org} (2012 to 2022, Theory $\rightarrow$ Algorithms and Complexity). Affiliations that were not listed on csrankings (such as industry labs) where ranked by the program chair. Then for any paper, we take the best rank among all affiliations associated with the authors of that paper.  The p-values associated with the reported correlations are computed via the permutation test (with 10,000 permutations).

\rev{We first discuss the {\bf change in overall merit scores:} 
\begin{itemize} 
\item We find that 7.1\% reviews changed their overall merit scores whereas 92.9\% left them unchanged (column 3 of Table~\ref{TabNumChanged}). 
\item We compute Kendall's tau b correlation between the change in overall merit and the rank of the associated paper. A positive correlation means that the overall merit for a paper with a better rank increased after the initial submission. We find that the correlation coefficient is $0.019$ and is not statistically significant ($p = 0.7189$). 
\item Among the 32 reviews that changed overall merit, the mean change was $-0.4$ and median change was $-1$.  (Kendall's tau b correlation coefficient when 
 restricted to these 32 changed reviews is $0.09$.)
\end{itemize} 
}

\rev{Next we discuss the {\bf change in reviewer expertise:}
\begin{itemize} 
\item We find that 3.8\% reviews changed their overall merit scores whereas 96.2\% left them unchanged (column 4 of Table~\ref{TabNumChanged}). 
\item We investigate if the self-reported reviewer expertise increased if the paper was from a better-ranked affiliation and the reviewer initially gave a high overall merit score, or if the paper was from a lower-ranked affiliation and the reviewer initially gave a low overall merit score. In what follows, this would be implied by a positive correlation coefficient and the opposite by a negative correlation coefficient. We compute Kendall's tau b correlation between the affiliation-rank of the paper and the quantity (updated expertise - old expertise)*(1 if overall merit=accept and -1 if overall merit=reject). We find that the correlation coefficient is $0.118$ and is statistically significant at a level of 0.05 ($p = 0.0265$). 
\item Among the 17 reviews that changed self-reported reviewer expertise, the mean change was $+0.47$ and median change was $+1$. (Kendall's tau b correlation coefficient when restricted to these 17 changed reviews is $0.46$.)
\end{itemize}
}

Also observe that among the aforementioned reviews that changed, a non-negligible fraction indicated that they knew the author identities, which suggests that either these reviews are changed for reasons not having to do with author identities, or that initially hiding the identities from the paper had some effect despite the reviewers' knowledge of author identities from elsewhere.

Finally, in 5 of the 315 reviews by external reviewers, the reviewer changed their answer about knowledge of author identities after initial submission. In two of these cases, they changed from `no idea' to `know', one changed from `know' to `no idea', one from `guess' to `no idea', and one from `no idea' to `guess'. None of the 90 reviews by program committee members changed the answer to the question on knowledge of author identities.

\begin{table}
\centering
\begin{tabular}{|l|c|c|c|}   
    \hline
     & Author identity & Change  in overall merit & Change in reviewer expertise \\\hline
     Program & Know &1/20 = 5.0\% & 1/20 = 5.0\%  \\
     committee & Guess&1/11 = 9.1\% & 0/11 = 0\%  \\
     member& No idea &4/59 = 6.8\% & 1/59 = 1.7\%  \\\hline
     External & Know &5/90 = 5.6\% & 6/90 = 6.7\%  \\
     reviewer & Guess&6/76 = 7.9\% &  4/76 = 5.2\%  \\
     & No Idea &15/195 = 7.7\% & 5/195 = 2.6\%  \\\hline
     Total &  & 32/451 = 7.1\% & 17/451 = 3.8\%  \\\hline
\end{tabular}
\caption{Number and percentage of reviews that changed after initial submission.\label{TabNumChanged}} 
\end{table}


\section{Main results 2: Survey}
\label{SecSurvey}

We conducted an anonymous, optional survey among the participants of the ITCS 2023 conference. The survey was sent out via email on November 2, 2022 (soon after the paper acceptance decisions were announced) and was open till November 15, 2022. We first present the contents of the survey  as shown to the participants, and then present aggregated responses. 

\subsection{Questionnaire presented to participants} 
In what follows, we present the survey questionnaire verbatim as presented to the participants. 

\noindent\rule{\textwidth}{1pt}
Our field has long debated whether to anonymize authors in the review process. The main reported benefit of anonymizing authors is that of reducing (biased) dependence on author identities in the review, as has been found in a number of experiments in other fields. A number of cons are also often discussed: 
\begin{itemize}
\item Challenges in conflict-of-interest detection
\item Reviewers want to use author identities for some part of their review (e.g., using track record to gain confidence in whacky ideas or complex proofs)
\item Reduces ambiguity w.r.t. prior literature (e.g., can know that a preprint was by the same authors)
\item Imperfectness of anonymization.
\end{itemize}

ITCS 2023 adopted a middle ground where reviewers initially did not see author identities, but could see after submitting their initial reviews, and could then modify their reviews. We would love to know your thoughts on this debate of anonymizing authors. When answering, please assume that conflict-of-interest detection is taken care of even if authors are anonymized (as done in many other conferences that adopt author anonymization).

~\\

\noindent I participated in ITCS 2023 as:
\begin{itemize}
 \renewcommand{\labelitemi}{$\square$}
    \item PC member
    \item External reviewer
    \item Author
\end{itemize}

\noindent Preferences regarding anonymizing authors:\plosfootnote{The ordering of the first four options was randomized to be either this order or reverse order.}
\begin{itemize}
 \renewcommand{\labelitemi}{$\square$}
    \item I prefer anonymizing author identities through the entire review process
    \item I prefer anonymizing author identities until the time for PC meetings/discussions
    \item I prefer anonymizing author identities until the reviewer submits their initial review
    \item I prefer not anonymizing author identities
    \item Other (please specify) $\_\_\_\_\_\_\_\_\_\_\_\_\_\_\_\_\_\_\_\_\_\_\_\_\_\_\_\_\_\_\_\_\_\_\_\_\_\_\_\_\_\_$
\end{itemize}

\noindent 
Please share the reasons for your preferences, as well as any other comments/experiences/opinions on this topic: $\_\_\_\_\_\_\_\_\_\_\_\_\_\_\_\_\_\_\_\_\_\_\_\_\_\_\_\_\_\_\_\_\_\_\_\_\_\_\_\_\_\_\_\_\_\_\_\_\_\_\_\_\_\_\_\_\_\_\_\_\_\_\_\_\_\_\_\_\_\_\_\_\_\_\_\_\_\_\_\_\_\_\_\_$

\noindent\rule{\textwidth}{1pt}


\subsection{Analysis of survey responses}
We received 200 responses to the questionnaire.  In this section, we summarize and analyze these responses. \rev{The full set of responses is available at~\cite[supporting information Data S1]{shah2022role}.}

\definecolor{colorA}{HTML}{003f5c}
\definecolor{colorD}{HTML}{8a61a5}
\definecolor{colorI}{HTML}{ef5675}
\definecolor{colorN}{HTML}{ffa600}
\definecolor{colorO}{HTML}{ffffff}

\newcommand{\patternA}{}
\newcommand{\patternD}{north east lines}
\newcommand{\patternI}{fivepointed stars}
\newcommand{\patternN}{bricks}
\newcommand{\patternO}{dots}

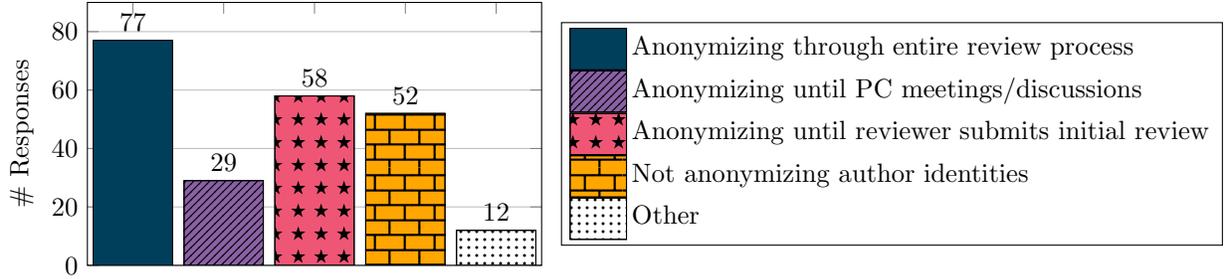
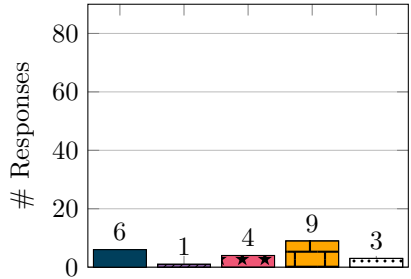
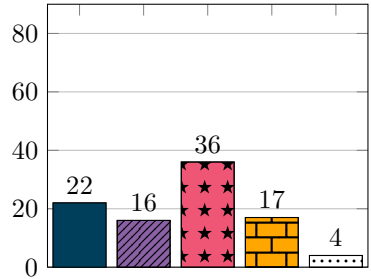
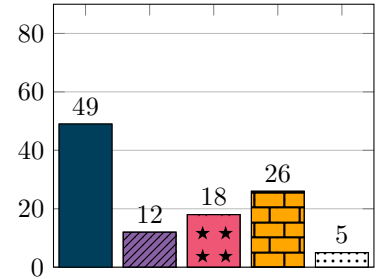
\begin{figure*}
\centering
\begin{subfigure}{\textwidth}
\begin{tikzpicture}
\begin{axis} [xtick={0, 1,2,3,4},
        width=3in,
        height=2in,
        ylabel = {\# Responses},
        ymin=0, ymax=90,
         nodes near coords, 
        xmin=-.5,xmax=4.5,
        xticklabels={,,,,},
        xticklabel style   = {align=center},
        every axis plot/.append style={
          ybar,
          bar width=30pt,
          bar shift=0pt,
          fill
        },
        ymajorgrids=true,
        legend style={at={(2.5,0.5)}, anchor=east,legend columns=1},
        legend image code/.code={
        \draw [#1] (1cm,-.3cm) rectangle (0.25cm,0.25cm); },
        legend cell align={left}
    ]
\addplot[fill = colorA, postaction={pattern=\patternA}] coordinates{
    (0,77)
};
\addplot[fill = colorD, postaction={pattern=\patternD}] coordinates{
    (1,29)
};
\addplot[fill = colorI, postaction={pattern=\patternI}] coordinates {
    (2,58)
};
\addplot [fill=colorN, postaction={pattern=\patternN}] coordinates {
    (3,52)
};
\addplot [fill=colorO, postaction={pattern=\patternO}] coordinates {
    (4,12)
};
\legend{Anonymizing through entire review process,Anonymizing until PC meetings/discussions,Anonymizing until reviewer submits initial review, Not anonymizing author identities, Other}
\end{axis}
\end{tikzpicture}
\caption{All responses (200 responses)\label{FigSurveyAll}}
\end{subfigure}\\~\\
\begin{subfigure}{.35\textwidth}
\begin{tikzpicture}
\begin{axis} [xtick={0, 1,2,3,4},
        width=2.3in,
        height=2in,
        ylabel = {\# Responses},
        ymin=0, ymax=90,
         nodes near coords, 
        xmin=-.5,xmax=4.5,
        xticklabels={,,,,},
        xticklabel style   = {align=center},
        every axis plot/.append style={
          ybar,
          bar width=20pt,
          bar shift=0pt,
          fill
        },
        ymajorgrids=true
    ]
\addplot[fill = colorA, postaction={pattern=\patternA}] coordinates{
    (0,6)
};
\addplot[fill = colorD, postaction={pattern=\patternD}] coordinates{
    (1,1)
};
\addplot[fill = colorI, postaction={pattern=\patternI}] coordinates {
    (2,4)
};
\addplot [fill=colorN, postaction={pattern=\patternN}] coordinates {
    (3,9)
};
\addplot [fill=colorO, postaction={pattern=\patternO}] coordinates {
    (4,3)
};
\end{axis}
\end{tikzpicture}
\caption{Program committee (PC) members (19 responses)\label{FigSurveyPC}}
\end{subfigure}
~
\begin{subfigure}{.3\textwidth}
\begin{tikzpicture}
\begin{axis} [xtick={0, 1,2,3,4},
        width=2.3in,
        height=2in,
        ymin=0, ymax=90,
         nodes near coords, 
        xmin=-.5,xmax=4.5,
        xticklabels={,,,,},
        xticklabel style   = {align=center},
        every axis plot/.append style={
          ybar,
          bar width=20pt,
          bar shift=0pt,
          fill
        },
        ymajorgrids=true
    ]
\addplot[fill = colorA, postaction={pattern=\patternA}] coordinates{
    (0,22)
};
\addplot[fill = colorD, postaction={pattern=\patternD}] coordinates{
    (1,16)
};
\addplot[fill = colorI, postaction={pattern=\patternI}] coordinates {
    (2,36)
};
\addplot [fill=colorN, postaction={pattern=\patternN}] coordinates {
    (3,17)
};
\addplot [fill=colorO, postaction={pattern=\patternO}] coordinates {
    (4,4)
};
\end{axis}
\end{tikzpicture}
\caption{External reviewers\\(84 responses)\label{FigSurveyExternal}}
\end{subfigure}
~
\begin{subfigure}{.3\textwidth}
\begin{tikzpicture}
\begin{axis} [xtick={0, 1,2,3,4},
        width=2.3in,
        height=2in,
        ymin=0, ymax=90,
         nodes near coords, 
        xmin=-.5,xmax=4.5,
        xticklabels={,,,,},
        xticklabel style   = {align=center},
        every axis plot/.append style={
          ybar,
          bar width=20pt,
          bar shift=0pt,
          fill
        },
        ymajorgrids=true
    ]
\addplot[fill = colorA, postaction={pattern=\patternA}] coordinates{
    (0,49)
};
\addplot[fill = colorD, postaction={pattern=\patternD}] coordinates{
    (1,12)
};
\addplot[fill = colorI, postaction={pattern=\patternI}] coordinates {
    (2,18)
};
\addplot [fill=colorN, postaction={pattern=\patternN}] coordinates {
    (3,26)
};
\addplot [fill=colorO, postaction={pattern=\patternO}] coordinates {
    (4,5)
};
\end{axis}
\end{tikzpicture}
\caption{Authors excluding external reviewers and PC (97 responses)\label{FigSurveyAuthors}}
\end{subfigure}
\caption{Answers given by respondents to the prompt ``Preferences regarding anonymizing authors.''\label{FigSurvey}} 
\end{figure*}

\subsubsection{Quantitative analysis}
\label{SecQuantitative}
We begin with an analysis of the responses to the quantitative question on the respondents' preferences about anonymizing authors. We report the results in Figure~\ref{FigSurvey}.\plosfootnote{The participants were allowed to choose more than one option. Here we report the total counts in terms of the number of respondents who chose any option, hence the sum of all options in Figure~\ref{FigSurvey} is greater than 200. An alternative approach would be as follows: If a participant chooses $k$ options, then count each chosen option as $\frac{1}{k}$. The result from this alternative approach is qualitatively similar to the current results, and hence we omit it for brevity.}  The key takeaway is that there is considerable support for anonymizing authors in at least some part of the review process, particularly from participants outside the program committee. 

Notice in Figure~\ref{FigSurveyAll} that twelve respondents chose the option `other.' These respondents also provided comments alongside this choice. The three PC members who selected `other' also indicated support for anonymization in some form, such as anonymization to reviewers but not to program committee members. Five of the twelve respondents who selected `other' and did not select any other option. Three of these five were supportive of anonymizing authors in peer review, but were concerned about challenges in detecting conflicts of interest (CoI). In absence of other ways for CoI detection, they supported anonymizing except for CoI detection. One of them strongly supported anonymization at least for external reviewers. One other respondent did not have an opinion. The remaining seven respondents who selected `other' also selected other options alongside, and are already counted in Figure~\ref{FigSurveyAll}. We discuss their text comments below along with all other general comments made by respondents.

\subsubsection{Free-text comments} 

We finally discuss the free-text comments provided by respondents. A total of 103 respondents provided free-text comments, and in what follows we summarize these comments. We saw in Figure~\ref{FigSurvey} that there were disagreements among respondents on the best policies for anonymizing author identities, and these disagreements are also reflected in the free-text comments.

We begin with {\bf comments that were common to a large number of respondents}.
\rev{
\begin{itemize}
    \item Respondents opine that revealing author identities can bias the reviews. 
    \item Respondents appreciated the initiative taken by ITCS 2023 in adopting a middle-ground approach. 
    \item Respondents complained about problems in avoiding conflicts-of-interest when assigning external reviewers. 
\end{itemize}

\noindent We now discuss the remaining comments, each of which was given {\bf by one or few respondents}.  We have broadly classified these comments in terms of their implications regarding author anonymization.  
\begin{itemize}
\item {\bf Problems with anonymization / benefits of knowing author identities}: 
    \begin{itemize}
                     \item Reviewers can get biasing information about authors, e.g., from preprints uploaded by authors or from the English writing style in the submission.

        \item Anonymization would create weird incentives around posting papers on preprint servers, where authors might delay posting papers to stay anonymous or post earlier to make their identities public, depending on what they think would benefit them.

        \item May subsequently lead to undesirable policies such as preprint and talk embargoes. 

        \item It can be challenging for authors to ensure that they do not inadvertently identify themselves.

        \item Anonymization makes rejecting a paper of a rival easier.
            
        \item Anonymization signals mistrust in the integrity of the community. 
        
        \item Authors may strongly criticize their own past work to bolster the current submission, or may make only incremental contributions without being noticed under anonymization. 
        
        \item Student authored papers should have lower bars, or can be shepherded instead of outright rejection.
        
        \item Some authors have a history of making errors or submitting bad papers, so knowing author identities can help assess correctness concerns.

        \item It is hard to assess quality in real time, and author reputation can serve as a good distinguisher, for instance, if the reviewer is not entirely familiar with the paper or if there are very long proofs.
                
        \item Reviewers should know who the authors are eventually, to give due credit if they learn something during the review process that they later use.

        \item If there are multiple closely related papers on a subject, it is important to understand to what extent these are distinct groups of authors.
        
        \item Need to specify anonymization policies for various situations, e.g., submitting a code repository.
\end{itemize}

\item {\bf Benefits of anonymizing}: 

\begin{itemize}
    \item Anonymization welcomes new authors and helps bring new ideas.
            
    \item Ideas and proofs should talk for themselves, and authors should explain new ideas well. Author name should not be a substitute. 
    
    \item It will force reviewers to read the paper carefully rather than judging the quality based on authors.
    
    \item It is for the papers in the grey area that author identity will bias decisions, and these papers are precisely the ones for which the discussion is critical.

    \item Revealing author identities may make the reviewer more careful if the reviewer knows that the identities will be known at the discussion phase. 
\end{itemize} 

\item {\bf Benefits of middle ground}:
\begin{itemize}
    \item Prevents initial bias but then helps reap benefits of not anonymizing, thereby achieving both goals. 
    
    \item People get curious and will look up the paper,  but they will be patient enough to wait until after the first review if they know they will get a chance to then see and potentially update their review. 
    
    \item Does not do any harm (presuming the chair can see the changes made).
\end{itemize}

\item {\bf Problems with middle ground}:
\begin{itemize}
    \item Puts back the bias into the process.
\end{itemize}

\item {\bf Suggestions}: 
\begin{itemize}
    \item The program committee members should be able to see the name in the online peer-review platform but not on the paper.
    
    \item Ask the reviewers to give two scores, one on the quality of the paper/result/novelty and another on confidence in the correctness/technical part. Among them, only the ``correctness confidence score" can be modified after seeing the authors' names after submitting an initial review.
      
    \item It should be the reviewer's choice as to whether they want author identities anonymized or not.
        
    \item Prohibiting arXiv and other external dissemination is not good, but there is no harm in anonymizing within the review process.

    \item Can have a two-tiered program committee where the senior members are aware of the author identities but do not actively participate in reviewing, but helps to resolve any issues with conflicts of interest.
        
    \item In the middle ground, program chairs should be able to see any change in the reviews once the reviewer sees the author identities.
\end{itemize}

\end{itemize}
}

\section{Discussion and limitations}
Our results suggest support for some form of anonymizing authors in peer review. Doing so can help mitigate biases pertaining to initial impressions about authors. In the literature, there are some stated benefits of not anonymizing authors, such as using author identities to gain more confidence in hard-to-verify mathematical proofs or wacky ideas, more scrutiny for papers by authors with a history of bad science, ensuring similar papers are from different authors, mitigating incremental work by same authors, etc. While these stated benefits are debated in the survey responses, if a venue wishes to realize them, the approach followed by ITCS 2023 may allow them to do so. Although this may result in bias appearing in the process after revelation of author identities, in ITCS 2023, we did not find much change in the overall merit scores after author identities were revealed. (Note that we reported being unable to check changes in review text or eventual discussions, leaving open the possibility of author identities playing a role there.) It is also sometimes stated that anonymizing authors is not useful as author identities may be guessed from the contents of the paper. However, we found that a majority of reviewers were unable to guess the authors' identities. 

Another stated benefit of knowing author identities is the ease of checking conflicts of interest. This was indeed a challenge at ITCS 2023, where several participants in the survey complained about problems in ensuring that papers are assigned to external reviewers who do not have conflicts of interest with the paper. Thus an anonymization of authors should be accompanied by an efficient and rigorous process to check conflicts of interest. One way to do so is to have an automated system to check for co-authorship and affiliation conflicts, as is done in machine learning conferences. An alternative ``manual'' option is to have a small set of volunteers who can check conflicts for any external reviewer that a program committee member wishes to invite for reviewing. A third possible solution, which leads to some reduction in anonymity, is to make author identities anonymous only to external reviewers but not program committee members. 

Note that importantly, the conference did not impose any restrictions on authors regarding posting their (non anonymous) papers elsewhere such as on preprint servers. Such policy choices are widely debated~\cite{rastogi2022arxiv} in some other research communities, where arguments in favor of such restrictions point to ensuring more thorough anonymization, whereas arguments against such restrictions include allowing free and open dissemination of research and furthermore of unintentionally biasing against the very people that it is supposed to protect~\cite[Chapter 7]{shah2022surveyextended}. The policies of ITCS 2023, as well as the analysis and discussion in the present paper pertain to the absence of such restrictions. 

This study also has several limitations that we discuss below.
\begin{itemize}
    \item As briefly mentioned above, we did not have access to any changes in the text of the reviews.\plosfootnote{Our preregistration included analysis of the change of text. However, we could not access the logs of the review text from the HotCRP conference management platform on which the peer-review process was conducted.} Author identities may also have played a role in the discussions between program committee members but we do not have logs of these discussions to draw any inference.
    \item One manner in which reviewers may use author identities is to gain confidence in their evaluation. We attempt to measure this, however, the review questionnaire asked reviewers to self report their expertise. The confidence and expertise of any reviewer may be related, but not identical. For instance, a self-report of expertise may pertain to the topic of the paper whereas a self-report of confidence may pertain to their evaluation of the paper. 
    \item The survey we conducted was anonymous, and hence comes with the usual caveats associated with anonymous surveys such as possible selection biases. We do stratify the responses by the respondents self-reported role in ITCS 2023 (program committee member, external reviewer, or author) but investigation of selection biases or stratification along other attributes is not possible. 

    \item The experiment was announced to reviewers prior to them starting to review, and they were told that the logs of the review revisions will be analyzed. It is thus possible that this information may have changed the reviewer's behavior (Hawthorne effect): \rev{ the reviewer may be hesitant to change their reviews after seeing the author identities thinking that a change in their review could be undesirably perceived as tied to knowledge of author identities.} 
\end{itemize} 
As for the last point above, we think that future conferences adopting such policies of partial anonymization should consider imparting transparency with respect to dependence on author identities, in which the program chairs, or the program committee, or even the authors can see the revisions of any review. 

All in all, we hope that this investigation will lead to more discussion and evidence-based policy design for an improved peer-review process.


\section*{Acknowledgments} 
We sincerely thank the ITCS 2023 program chair Yael Kalai for trying out this middle ground approach to author anonymization, for facilitating the analysis, and for  valuable inputs throughout the analysis. We also thank Pravesh Kothari for very helpful discussions. We are grateful to all the program committee members and reviewers of ITCS 2023 for their efforts in the review process, and all the respondents of our survey for sharing their opinions and suggestions. 

\let\OLDthebibliography\thebibliography
\renewcommand\thebibliography[1]{
  \OLDthebibliography{#1}
  \setlength{\parskip}{5pt}
  \setlength{\itemsep}{0pt plus 0.3ex}
}
\printbibliography

\end{document}